\documentclass{article}
\PassOptionsToPackage{square,comma,numbers,sort&compress}{natbib}

 \usepackage{amsmath}

 \usepackage[main, preprint]{neurips_2025}

\usepackage[utf8]{inputenc} 
\usepackage[T1]{fontenc}    
\usepackage{hyperref}       
\usepackage{url}            
\usepackage{booktabs}       
\usepackage{amsfonts}       
\usepackage{nicefrac}       
\usepackage{microtype}      
\usepackage{xcolor}         
\usepackage{graphicx} 
\usepackage{booktabs}
\usepackage{multirow}

\title{MoE-TTS: Enhancing Out-of-Domain Text Understanding for Description-based TTS via Mixture-of-Experts}

\author{%
  {Heyang Xue, Xuchen Song\thanks{Corresponding author} , Yu Tang, Jianyu Chen, Yanru Chen, Yang Li, Yahui Zhou}\\
  Kunlun Inc.\\
  \texttt{\{heyang.xue, xuchen.song\}@kunlun-inc.com}\\
}

\begin{document}

\maketitle

\begin{abstract}
Description-based text-to-speech~(TTS) models exhibit strong performance on in-domain text descriptions, i.e., those encountered during training. However, in real-world applications, the diverse range of user-generated descriptions inevitably introduces numerous out-of-domain inputs that challenge the text understanding capabilities of these systems. To address this issue, we propose MoE-TTS, a description-based TTS model designed to enhance the understanding of out-of-domain text descriptions. MoE-TTS employs a modality-based mixture-of-experts~(MoE) approach to augment a pre-trained textual large language model~(LLM) with a set of specialized weights adapted to the speech modality while maintaining the original LLM frozen during training. This approach allows MoE-TTS to effectively leverage the pre-trained knowledge and text understanding abilities of textual LLMs.  Our experimental results indicate that: first, even the most advanced closed-source commercial products can be challenged by carefully designed out-of-domain description test sets; second, MoE-TTS achieves superior performance in generating speech that more accurately reflects the descriptions. We encourage readers to listen to the demos at \url{https://welkinyang.github.io/MoE-TTS/}.
\end{abstract}

\section{Introduction}
In recent years, description-based text-to-speech~(TTS) technology has been adopted in industrial applications~\cite{ElevenLabs, MiniMax}, enabling users to precisely control the speaker and style characteristics of synthesized speech through natural language text descriptions (e.g., "clear, youthful voice with a magnetic tone"). This interaction method significantly lowers the barrier for speech customization and holds great potential in areas such as virtual assistants and audio content creation. Concurrently, substantial research~\cite{DBLP:conf/icassp/GuoLWZT23, DBLP:conf/icassp/ShimizuYKSDKT24, DBLP:conf/icassp/JiZ00CDHZ24, DBLP:conf/iclr/LengGSJ0LLYZS0024, DBLP:conf/interspeech/KawamuraYSHT24, DBLP:conf/mm/Jin0W0ZZQ024, DBLP:journals/corr/abs-2402-01912, DBLP:journals/corr/abs-2503-04713} progress has been made to generate speech that better aligns with these descriptions. Among them, numerous studies~\cite{DBLP:conf/icassp/JiZ00CDHZ24, DBLP:conf/icassp/ShimizuYKSDKT24,DBLP:conf/mm/Jin0W0ZZQ024,  DBLP:conf/interspeech/KawamuraYSHT24, DBLP:journals/corr/abs-2402-01912,  DBLP:journals/corr/abs-2503-04713} have contributed dedicated, open-source datasets to advance description-based TTS. The natural language descriptions within these datasets often originate from a finite set of predefined tags representing speaker and style attributes. While dataset designers utilize large language models~(LLMs)~\cite{DBLP:journals/corr/abs-2303-08774, DBLP:journals/corr/abs-2307-09288} and carefully engineered prompts to structure these tags into natural language and enhance diversity, the resulting descriptions~(referred to as in-domain descriptions) remain fundamentally constrained by the underlying tag space. In stark contrast, real-world user-generated descriptions exhibit immense diversity, inevitably exceeding the scope of the training data. These out-of-domain descriptions pose a significant challenge to the text understanding capabilities of models trained solely on in-domain data, limiting their practical robustness. 

Existing approaches to text understanding in description-based TTS fall into three main categories. The first approach involves training encoders for textual descriptions jointly with the TTS model itself~\cite{DBLP:conf/icassp/ShimizuYKSDKT24, DBLP:conf/iclr/LengGSJ0LLYZS0024}. Although this method is straightforward, it relies exclusively on in-domain descriptions for learning, which inherently limits its potential for broader generalization. The second approach~\cite{DBLP:conf/icassp/GuoLWZT23, DBLP:journals/corr/abs-2402-01912} incorporates pre-trained textual encoders, such as T5~\cite{DBLP:journals/jmlr/RaffelSRLNMZLL20}, to leverage extensive pre-trained linguistic knowledge. Although this strategy improves generalizability to some extent, it is constrained by the capacity of the encoder to handle complex linguistic phenomena. The last category~\cite{DBLP:journals/corr/abs-2412-10117, DBLP:journals/corr/abs-2504-12867} uses a pre-trained textual LLM to initialize the backbone network without any additional encoders, thereby providing a more natural mechanism for processing natural language descriptions. Notably, textual LLMs (e.g. Qwen~\cite{DBLP:journals/corr/abs-2505-09388}) offer not only richer pre-trained knowledge but also more advanced natural language understanding capabilities, enabling them to decode nuanced descriptions. Despite the promising potential of this third approach for out-of-domain understanding, updating all model parameters within a modal coupled framework can lead to catastrophic forgetting~\cite{DBLP:journals/corr/abs-2309-10313}, which undermines the retention of pre-trained knowledge and consequently impairs text understanding. To address this challenge, the present work investigates improved strategies for leveraging textual LLMs to achieve robust understanding of out-of-domain descriptions.

Inspired by the successful application of the Mixture-of-Experts~(MoE) paradigm in multimodal large language models~(MLLMs)~\cite{DBLP:conf/cvpr/LuoYDWLD0Z25, DBLP:journals/corr/abs-2502-06788, DBLP:conf/nips/BaoW0LMASPW22, DBLP:journals/corr/abs-2208-10442}, we propose MoE-TTS. Our model enhances out-of-domain description understanding within a description-based TTS framework. Specifically, MoE-TTS integrates a set of learnable speech-specific parameters (acting as speech-modality experts) seamlessly into a pre-trained textual LLM via a modality-based MoE approach. Crucially, the original LLM parameters remain frozen throughout training, preserving its potent pre-trained knowledge and text understanding capabilities. Furthermore, we meticulously construct an out-of-domain description test set using diverse linguistic strategies specifically designed to rigorously evaluate the generalization performance of MoE-TTS on challenging unseen descriptions. In our experiments, we compared MoE-TTS with state-of-the-art models from commercial entitles, such as: ElevenLabs~\cite{ElevenLabs} and MiniMax~\cite{MiniMax}. The results suggest that even when trained solely on open-source datasets, MoE-TTS still demonstrates superior performance for both in-domain and out-of-domain descriptions. Our key contributions can be summarized as follows:
\begin{itemize}
    \item We are the first to focus on the performance of description-based TTS with out-of-domain descriptions. This approach helps bridge the gap between description-based TTS research and real-world applications. 
    \item To the best of our knowledge, MoE-TTS is the first work to apply mixture-of-experts techniques to enhance TTS models by leveraging the pre-trained knowledge and text understanding capabilities of textual LLM.
    \item The experimental results demonstrate that our design surpasses state-of-the-art commercial models in generating speech that more aligned with the descriptions.
    
\end{itemize}

\section{Related Work}
\label{gen_inst}
\subsection{Description-based TTS}
PromptTTS~\cite{DBLP:conf/icassp/GuoLWZT23} pioneered description-based TTS by creating a natural language dataset derived from five style tags using SimBERT~\cite{simbert}. Its architecture utilized BERT~\cite{DBLP:conf/naacl/DevlinCLT19} as the text understanding module and FastSpeech~\cite{DBLP:conf/nips/RenRTQZZL19} as the speech generation module. While PromptTTS demonstrated the ability to control speaker and style characteristics through natural language descriptions, it exhibited notable limitations. Specifically, its tag-based system was oversimplified: the limited number of tags provided a restricted descriptive vocabulary for each tag.

To address this data limitation, subsequent research significantly expanded the tagging framework. SpeechCraft~\cite{DBLP:conf/mm/Jin0W0ZZQ024} increased the number of tags to eight, while ParaspeechCaps~\cite{DBLP:journals/corr/abs-2503-04713} extended it substantially to 59 tags, enriching the descriptive vocabulary associated with each tag. Furthermore, later studies replaced SimBERT with powerful, commercially available large language models~(LLMs), such as GPT-3.5 Turbo in TextrolSpeech~\cite{DBLP:conf/icassp/JiZ00CDHZ24} and GPT-4 in ParaspeechCaps~\cite{DBLP:conf/iclr/LengGSJ0LLYZS0024}, to generate more natural and diverse descriptions from the tags. EmoVoice~\cite{DBLP:journals/corr/abs-2504-12867} constructed a high-quality emotion dataset featuring expressive speech and fine-grained emotion tags with natural language descriptions.  Concurrently, advancements in language models-based TTS~\cite{DBLP:journals/corr/abs-2301-02111, DBLP:journals/corr/abs-2412-10117, DBLP:journals/corr/abs-2406-02430, DBLP:journals/corr/abs-2505-07916} led to their replacement of FastSpeech as generation modules~\cite{DBLP:conf/icassp/JiZ00CDHZ24, DBLP:journals/corr/abs-2402-01912, DBLP:journals/corr/abs-2504-12867}. Models like Parler-TTS~\cite{DBLP:journals/corr/abs-2402-01912} further enhance text processing by leveraging pre-trained T5 within the understanding module. 

Collectively, these works and the open-source datasets they introduced—have established a critical foundation for ongoing research in description-based TTS systems.

\subsection{Mixture-of-Experts}
Recently, the MoE paradigm has been widely adopted in MLLMs~\cite{DBLP:conf/cvpr/LuoYDWLD0Z25, DBLP:journals/corr/abs-2502-06788, DBLP:conf/nips/BaoW0LMASPW22, DBLP:journals/corr/abs-2208-10442, DBLP:conf/emnlp/ShenYLDKH23}. These models benefit from the ability of MoE approaches to decouple parameter spaces across different modalities, thereby avoiding the representational interference often caused by fully shared parameters. This facilitates more effective training and model scaling~\cite{DBLP:conf/emnlp/ShenYLDKH23}.

Notably, Mono-InternVL~\cite{DBLP:conf/cvpr/LuoYDWLD0Z25} leverages MoE by embedding a visual parameter space into a frozen, pre-trained textual LLM and employs a static routing strategy to assign experts to corresponding tokens. This design preserves the pre-trained knowledge and text understanding capabilities of the LLM while facilitating the learning of visual representations. Furthermore, EVEv2~\cite{DBLP:journals/corr/abs-2502-06788} discovered the shortcomings caused by interference between modalities in multimodal learning through a series of careful experiments. To address this issue, EVEv2 incorporates modality-specific weights into critical Transformer components achieving finer-grained modality separation.

Our work is inspired by these innovative approaches to modality integration and parameter decoupling.

\subsection{Text Understanding Ability}
Previous studies~\cite{DBLP:journals/corr/abs-2502-04128} have established that robust text understanding is critical for the performance of TTS models, influencing naturalness, content accuracy, and expressiveness. Consequently, leveraging the inherent text-processing capabilities of textual LLMs presents a natural approach. For example, CosyVoice2~\cite{DBLP:journals/corr/abs-2412-10117} initializes its TTS model with Qwen2.5-0.5B~\cite{DBLP:journals/corr/abs-2501-15383}, and Llasa~\cite{DBLP:journals/corr/abs-2502-04128} inherits initialization parameters and training paradigms from LLaMA3~\cite{DBLP:journals/corr/abs-2407-21783}.

However, catastrophic forgetting~\cite{DBLP:journals/corr/abs-2309-10313} during training hinders the full utilization of the text capabilities of LLMs when relying solely on initialization. Orpheus-TTS~\cite{Orpheus-TTS} attempts to mitigate this issue by incorporating text data during pre-training to preserve the capabilities of LLaMA. Nevertheless, this strategy introduces significant computational overhead due to the complexity of the data. In addition, discrepancies in the data distribution and scale of the incorporated text and the original LLM training corpus can compromise their linguistic competence.

In this paper, MoE-TTS addresses the challenge of effectively leveraging the pre-trained knowledge and text understanding capabilities of textual LLMs within TTS frameworks. While primarily focused on improving text description understanding, MoE-TTS provides a practical solution to this critical issue by effectively mitigating the limitations related to catastrophic forgetting and the complexities of multi-modal data integration.

\begin{figure*}[t]
    \centering
    \includegraphics[width=\linewidth]{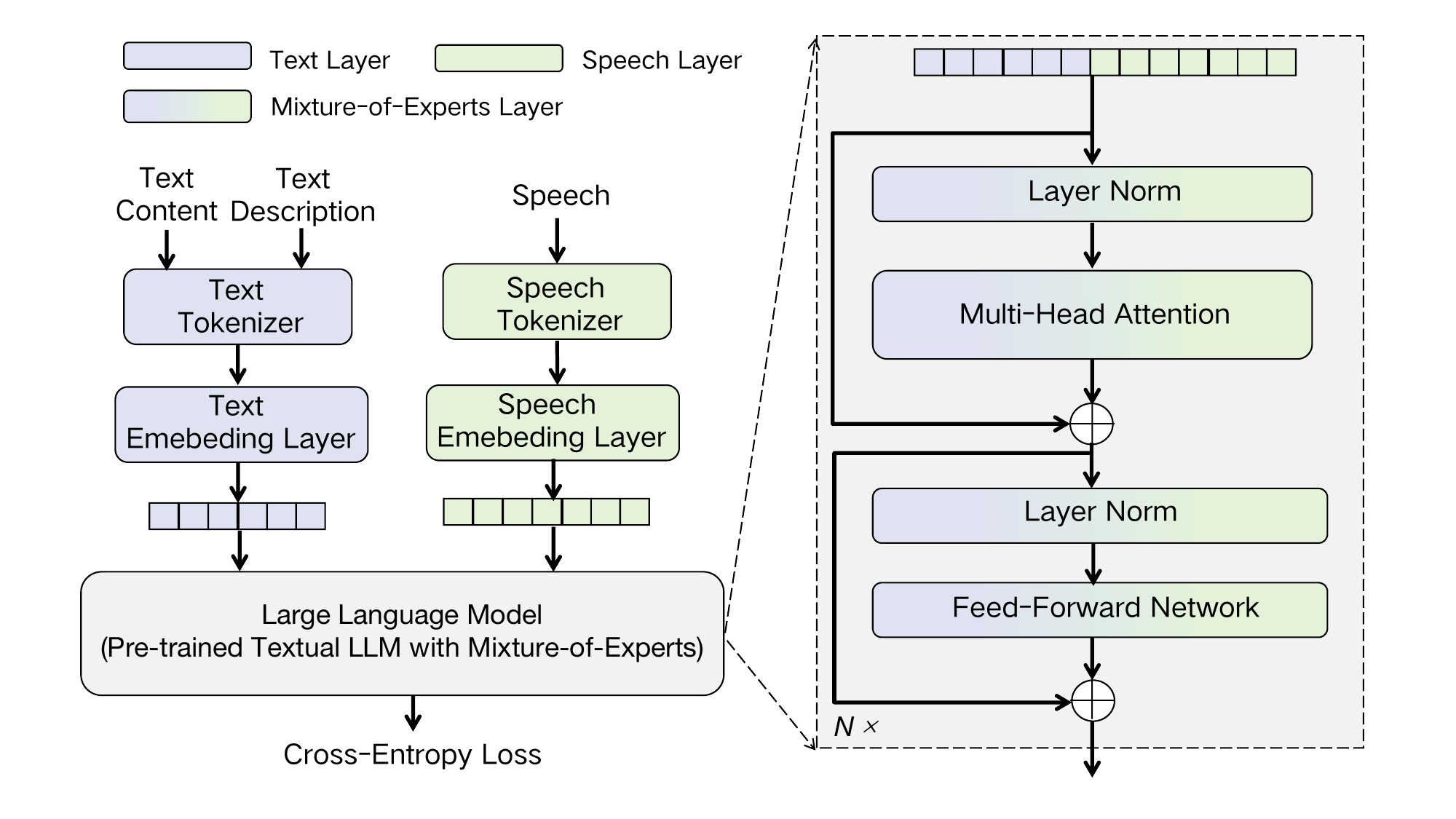}
    {

    \vspace{-1em}
    \caption{
    Overview of MoE-TTS. MoE-TTS is initialized from a pre-trained textual LLM and transforms key components of the original Transformer blocks into mixture-of-expert layers. The original weights function as text experts, while the newly incorporated weights serve as speech experts.
    \label{fig:framework}
    }
    }
\end{figure*}

\begin{table}[t]

 \caption{Examples of differences in wording between in-domain descriptions and out-of-domain descriptions for different voice attributes. The underline indicates the description corresponding to the specific voice attribute.}
 \label{tab:diff}
\resizebox{\columnwidth}{!}{
\begin{tabular}{@{}ccc@{}}
\toprule
Voice Attributes & In-domain Descriptions                                                                                                                                                                                                                                                           & Out-of-domain Descriptions                                                                                                                                                                                                                                                                         \\ \midrule
Pitch\&Speed     & \begin{tabular}[c]{@{}c@{}}A female speaker delivers a monoton\\ e speech with \underline{a slightly high-pitch voice} and \underline{moderate speed}.\end{tabular}                                                                                                                                      & \begin{tabular}[c]{@{}c@{}}\underline{Talking Taser}, Female, 20-35, chipmunk-on-jet-fuel energy. \\ \underline{Words fire like a machine gun}, punctuated by dolphin-like yips. \\ Occasionally short-circuits into gibberish.\end{tabular}                                                                               \\ \midrule
Gender           & \begin{tabular}[c]{@{}c@{}}\underline{A female speaker} with an Indian accent delivers a speech at a slow pace\\  in a slightly clean environment. \\ \underline{Her voice} is characterized as flowing and high-pitched.\end{tabular}                                                                   & \begin{tabular}[c]{@{}c@{}}Middle-aged puppet marriage counselor, unusually calm, \\ with a silky smooth yet hollow \underline{contralto voice}, \\ speaking at a suffocating pace.\end{tabular}                                                                                                               \\ \midrule
Accent           & \begin{tabular}[c]{@{}c@{}}In a flowing, slightly noisy environment, a male speaker delivers \\ his speech with a medium-pitched, \\ deep voice at a measured speed, displaying an \underline{American accent}.\end{tabular}                                                                 & \begin{tabular}[c]{@{}c@{}}\underline{US actor} with \underline{a New York accent}, versatile, articulate, \\ with a dynamic pace, full of charm and charisma, \\ attracting the attention of the audience.\end{tabular}                                                                                                   \\ \midrule
Speaking Style   & \begin{tabular}[c]{@{}c@{}}In the context of News and Politics, a calm adult male with a high pitch and \\ low voice demonstrates \underline{confidence and composure} as he utters.\end{tabular}                                                                                            & \underline{Young US rapper, speaks like a rap}, angry tone, fast and concise.                                                                                                                                                                                                                                  \\ \midrule
Timbre           & \begin{tabular}[c]{@{}c@{}}A man speaks fast with normal pitch and high energy.\\  Descriptions of the speaker’s vocal style are \underline{masculine},\\ \underline{adult-like},\underline{thin},\underline{slightly muffled,fluent},\\ \underline{cool,intellectual,calm,slightly friendly,reassuring,lively,slightly strict}\end{tabular} & \begin{tabular}[c]{@{}c@{}}Dragon Chess Master, male, 90 years old, exhales smoke when thinking for a long time, \\ \underline{with a hoarse bronze chime tone.} \\ \underline{Every sentence feels like striking a bell buried in volcanic ash for thousands of years,} \\ \underline{with a burning tremor at the end.}\end{tabular} \\ \midrule
Volume           & \begin{tabular}[c]{@{}c@{}}Engaging in a discussion about Science and Technology,\\  a sad young male with low pitch and \underline{normal volume} spoke at a normal pace, \\ referring to a previous conversation about Goldman.\end{tabular}                                               & \begin{tabular}[c]{@{}c@{}}9-year-old girl dragon tamer in the animation, excited tone, \\ \underline{occasional attempts to roar.}\end{tabular}                                                                                                                                       \\ \bottomrule
\end{tabular}}
\end{table}

\section{MoE-TTS}
\label{headings}

\subsection{Basic Task}

Our core idea is to fully leverage the pre-trained knowledge of textual LLMs to enhance the understanding of out-of-domain descriptions. Following this principle, we first construct the entire description-based TTS system from a pre-trained transformer-based textual LLM with a textual tokenizer~(e.g., Qwen3) and adhering to its next-token prediction training objective. To enable the generation of speech tokens, we then incorporate speech tokens into the vocabulary and insert the newly initialized token embeddings into the original LLM. This approach allows us to train description-based TTS models within a fully textual LLM paradigm.  Specifically, given the conditional text input \( T \in \mathbb{Z}^n \), which consists of the text transcription and text description, we have the corresponding speech output \( Y \in \mathbb{R}^m \). We then convert \( T \) and \( Y \) into discrete text tokens \( t \in \mathbb{R}^n \) and speech tokens \( y \in \mathbb{R}^{m'} \) through text and speech tokenizers, respectively. In this work, MoE-TTS addresses the following problem:
\begin{equation}
    \begin{aligned}
        p(y_{1:m'} \mid t_{1:n};\theta, \theta_{s}) = p(y_{1} \mid t_{1:n};\theta, \theta_{s}) \prod_{i=2}^{m'} p(y_{m'} \mid t_{1:n}, y_{1:m'-1};\theta, \theta_{s}).
    \end{aligned} 
    \label{eq_arch}
\end{equation}
where \(\theta\) denotes the parameters of the pre-trained LLM, and \(\theta_{s}\) represents the newly added parameters for the speech modality.

\subsection{Modality-based Mixture-of-Experts}

Using a textual LLM as the base model leverages its pre-trained knowledge and text understanding capabilities. However, updating LLM parameters within a modal coupled framework during training inevitably leads to catastrophic forgetting~\cite{DBLP:conf/cvpr/LuoYDWLD0Z25}, which hinders the effective utilization of that pre-trained knowledge and text understanding capabilities. To address this issue, we introduce a practical MoE approach. As illustrated in Figure~\ref{fig:framework}, MoE-TTS integrates a set of speech-modality parameter spaces, acting as speech experts, into the pre-trained LLM. It employs a modality-specific routing strategy~\cite{DBLP:conf/cvpr/LuoYDWLD0Z25} to assign text and speech experts to their corresponding tokens, resulting in modality-aware MoE layers. During training, only the weights of the speech-modality experts are updated, while the original LLM parameters remain frozen. This approach ensures that the frozen parameters retain the pre-trained knowledge, thereby preventing catastrophic forgetting. Moreover, these frozen parameters enhance generalization on out-of-domain descriptions during inference by preserving the powerful text understanding capabilities of the original textual LLM. To further reduce interference between modalities, we follow~\cite{DBLP:journals/corr/abs-2502-06788} by converting critical components within transformer blocks into modality-aware MoE layers, including multi-head attention (ATTN), feed-forward networks (FFN), and layer normalization (LN).

Specifically, given text tokens $t \in \mathbb{R}^n$ and speech tokens $y \in \mathbb{R}^{m'}$ as inputs, we first obtain token embeddings:
\begin{equation}
    \begin{aligned}
        e_{t} = \text{Embedding}(t), \quad e_{y} = \text{Embedding}(y)
    \end{aligned} 
\end{equation}
yielding $e_{t} \in \mathbb{R}^{n \times d}$ and $e_{y} \in \mathbb{R}^{m' \times d}$. The combined input token embedding for transformer blocks is then $e = \text{Concat}(e_{t}, e_{y}) \in \mathbb{R}^{(n+m') \times d}$. With the MoE approach integrated, each transformer block layer operates as follows:
\begin{equation}
    \begin{aligned}
        e^{l'} &= e^{l-1} + \text{ATTN}(\text{MoE\_LN}(e^{l-1})), \\
        e^{l} &= e^{l'} + \text{MoE\_FFN}(\text{MoE\_LN}(e^{l'})). \\
        \text{ATTN}(e) &= \text{MoE\_{O}}\left( \text{softmax}\left( \frac{\text{MoE\_Q}(e) \cdot \text{MoE\_K}(e)^{\text{T}}}{\sqrt{d_{k}}} \right) \cdot \text{MoE\_V}(e) \right).
    \end{aligned} 
\end{equation}
All MoE layers prefixed with '$\text{MoE}\_$' are defined conditionally based on token modality:
\begin{equation}
    \text{MoE}\_\text{Layer}(e_{i}) =  
    \left\{
        \begin{aligned}
            &\text{Layer}_{t}(e_{i}), \quad \text{if} \enspace e_{i} \in e_{t}, \\
            &\text{Layer}_{y}(e_{i}), \quad \text{if} \enspace e_{i} \in e_{y}. \\
        \end{aligned} 
    \right.
\end{equation}
where \( e_{i} \) denotes the \( i \)-th element of the token embeddings \( e \). \(\text{Layer}_{t}\) and \(\text{Layer}_{y}\) represent the text expert and speech expert components of the MoE layers, respectively. The speech expert is initialized using the parameters of the text expert, ensuring that both layers possess an identical architecture and parameter count.

\subsection{Acoustic Modeling}
Since the central concept of MoE-TTS focuses on leveraging the text understanding capabilities of pre-trained textual LLMs, the model does not impose any restrictions on the choice of discrete speech representations. Available options include:
\begin{itemize}
    \item Semantic tokens derived from self-supervised learning (SSL) encoders~\cite{DBLP:journals/taslp/HsuBTLSM21, DBLP:conf/icml/ChiuQZYW22} using vector quantization methods~\cite{DBLP:conf/nips/OordVK17, DBLP:journals/taslp/ZeghidourLOST22}
    \item Acoustic tokens generated by neural codecs~\cite{DBLP:journals/taslp/ZeghidourLOST22, DBLP:conf/nips/KumarSLKK23}
    \item Hybrid representations combining both~\cite{DBLP:journals/corr/abs-2502-04128, DBLP:journals/corr/abs-2503-01710}
\end{itemize}
In this work, we utilize the speech tokenizer from CosyVoice2 for waveform tokenization. To reconstruct the waveform from predicted speech tokens, we employ a widely adopted multi-stage approach~\cite{DBLP:journals/corr/abs-2412-10117, DBLP:journals/corr/abs-2406-02430, DBLP:journals/corr/abs-2505-07916}. Specifically, a diffusion model~\cite{DBLP:conf/nips/KarrasAAL22} transforms discrete speech tokens predicted by the LLM into Gaussian latent representations. During training, these Gaussian latent representations are extracted using a pre-trained VAEGAN~\cite{DBLP:conf/icassp/EvansPCZTP25}. During inference, the decoder component of the VAEGAN converts these latent representations back into the final waveform.

\section{Experiments}
\subsection{Experimental Setup}
\paragraph{Model Setup} In this work, we utilize the open-sourced Qwen3-4B model~\cite{DBLP:journals/corr/abs-2505-09388} as the foundational textual LLM to leverage its powerful pre-training knowledge and text understanding capabilities. After implementing the previously described MoE approach, we scaled MoE-TTS to 8 billion parameters. For speech tokenization, we employ the speech tokenizer component of CosyVoice2 operating at a frame rate of 25 Hz to convert waveform signals into discrete speech tokens. These 6,561 distinct speech tokens were subsequently incorporated into the vocabulary of the Qwen3 model. Regarding the diffusion model implementation, we adopt the Elucidated Diffusion Models (EDM) framework~\cite{DBLP:conf/nips/KarrasAAL22}—a widely recognized approach in image and speech generation. Specifically, we implement the Diffusion Transformer (DiT) architecture~\cite{DBLP:conf/eccv/MaGABVX24} from Stable Audio~\cite{DBLP:conf/icassp/EvansPCZTP25}, configuring it with 0.7 billion parameters. Additionally, we integrate the VAEGAN component from Stable Audio, preserving the same architectural configuration, training procedures, and inference pipelines. This component processes latent representations at 50 Hz with 32-dimensional features.

\paragraph{Training Details} 
During training, we strictly adhered to the original chat template format of Qwen3. This involved utilizing system prompts, text descriptions, and text transcripts as user messages, while speech tokens were used as assistant messages. The template contains a thinking block designated for Chain-of-Thought~(CoT) content, which we left empty for simplicity.  Due to the limited scale of open-source description-based TTS datasets, we divided the LLM training within MoE-TTS into two phases:  
1) Pre-training phase: Focused on developing text-to-speech capabilities using pure TTS datasets devoid of text descriptions.  
2) Fine-tuning phase: Leveraged description-based datasets to enhance understanding of text descriptions and corresponding speech generation, initializing from the parameters obtained in phase one.  Both phases employed the AdamW optimizer with betas of [0.9, 0.98], a learning rate of \(3 \times 10^{-4}\), a cosine learning rate scheduler, and a warm-up ratio of 0.08, training for one epoch. Importantly, only the speech-modality parameters were updated during training, while the original parameters~(i.e., text-modality experts) remained frozen. Regarding the acoustic modeling modules (EDM and VAEGAN), we trained them exclusively on TTS datasets without any fine-tuning.

\paragraph{Data Preparation} To facilitate reproducibility, we exclusively utilized open-source corpora. For the LLM pre-training phase, we employed two TTS datasets: VoxBox~\cite{DBLP:journals/corr/abs-2503-01710} and Emilia-YODAS~\cite{DBLP:journals/corr/abs-2501-15907}. The fine-tuning phase incorporated multiple description-based TTS datasets: TextrolSpeech~\cite{DBLP:conf/icassp/JiZ00CDHZ24}, SpeechCraft~\cite{DBLP:conf/mm/Jin0W0ZZQ024}, Parler-TTS~\cite{DBLP:journals/corr/abs-2402-01912}, LibriTTS-P~\cite{DBLP:conf/interspeech/KawamuraYSHT24}, and ParaspeechCaps~\cite{DBLP:journals/corr/abs-2503-04713}. All speech samples were resampled to 16kHz to align with the speech tokenizer requirements during LLM training. For VAEGAN training, samples were resampled to 48kHz to ensure high-fidelity waveform reconstruction.  To evaluate the effectiveness of MoE-TTS, we first constructed two specific test sets: an in-domain description test set and an out-of-domain description test set. Each test sample includes a text description and the text to be synthesized. Specifically, the in-domain test set contains 20 test samples, while the out-of-domain test set contains 40 test samples. As the name suggests, the text descriptions in the in-domain description set do not exceed the scope of the training data. In contrast, the descriptions in the out-of-domain test set were constructed using a series of linguistic strategies to ensure they differ significantly from the training data. Specifically, we employed metaphors, analogies, implications, and paraphrases to implicitly convey the speaker and style voice attributes contained in the descriptions, such as gender, age, pitch, speed, speaking style, and emotion. Table~\ref{tab:diff} provides examples illustrating the differences between in-domain and out-of-domain descriptions when describing various voice attributes.

\begin{table}[t]

 \caption{We compare MoE-TTS with the most leading commercial products.}
 \label{tab:mos}
\resizebox{\columnwidth}{!}{\begin{tabular}{@{}cccccccc@{}}
\toprule
Test Sets                                   & Models     & SQ                  & WSSA                & PA                  & SEA                 & OA                  & OS         \\ \midrule
\multirow{3}{*}{In-domain Descriptions}     & MoE-TTS    & 4.12±0.082 & 4.06±0.108          & 4.23±0.106          & \textbf{4.06±0.093} & \textbf{3.61±0.132} & 3.82±0.081 \\
                                            & ElevenLabs & 4.06±0.084          & \textbf{4.27±0.084} & \textbf{4.45±0.094} & 3.85±0.098          & 3.26±0.142          & 3.77±0.070 \\
                                            & MiniMax    & \textbf{4.24±0.082}          & 4.11±0.103          & 4.40±0.094          & 3.87±0.098          & 3.46±0.132          & \textbf{3.83±0.081} \\ \midrule
\multirow{3}{*}{Out-of-domain Descriptions} & MoE-TTS    & 3.84±0.075          & 4.30±0.070          & 4.57±0.052          & \textbf{4.02±0.063} & \textbf{3.75±0.097} & \textbf{3.79±0.054} \\
                                            & ElevenLabs & 4.02±0.071          & \textbf{4.35±0.069} & \textbf{4.62±0.051} & 3.89±0.080          & 3.39±0.101          & 3.73±0.059 \\
                                            & MiniMax    & \textbf{4.25±0.062} & 4.28±0.066          & 4.56±0.060          & 3.87±0.072          & 3.30±0.091          & 3.70±0.063 \\ \bottomrule
\end{tabular}}
\end{table}

\subsection{Evaluations}

\paragraph{Evaluation Metrics} A well-designed description-based TTS model should meet two core requirements for effective evaluation:  1) The synthesized speech content must accurately reflect the target text transcription.
2) Speaker characteristics and style characteristics in synthesized speech must closely match the text description. To comprehensively assess model performance against these criteria, we designed a detailed subjective evaluation. Twenty-one professional speech evaluators participated, each rating synthesized samples on a 5-point scale across six dimensions: speech quality~(SQ), word and sentence segmentation accuracy~(WSSA), pronunciation accuracy~(PA), stylistic expressiveness alignment~(SEA), overall alignment~(OA) and overall score~(OS). Here, alignment quantifies conformity with the text descriptions. The dimensions SQ, WSSA, and PA evaluate requirement 1 (content accuracy), while SEA and OA measure requirement 2 (characteristic matching). The OA score represents overall performance considering all factors.  We computed Mean Opinion Scores (MOS) for each dimension by aggregating all evaluator ratings. Although alignment scores indicate overall conformity between description and speech, they lack detailed insights into contributing factors. Therefore, evaluators also verified attribute-level matches for gender, age, accent, and speaking speed. Each test sample underwent attribute verification by seven evaluators. Results were aggregated to calculate accuracy metrics per attribute, enabling a granular analysis of the factors influencing alignment scores.

\begin{figure*}[t]
    \centering
    \includegraphics[width=\columnwidth]{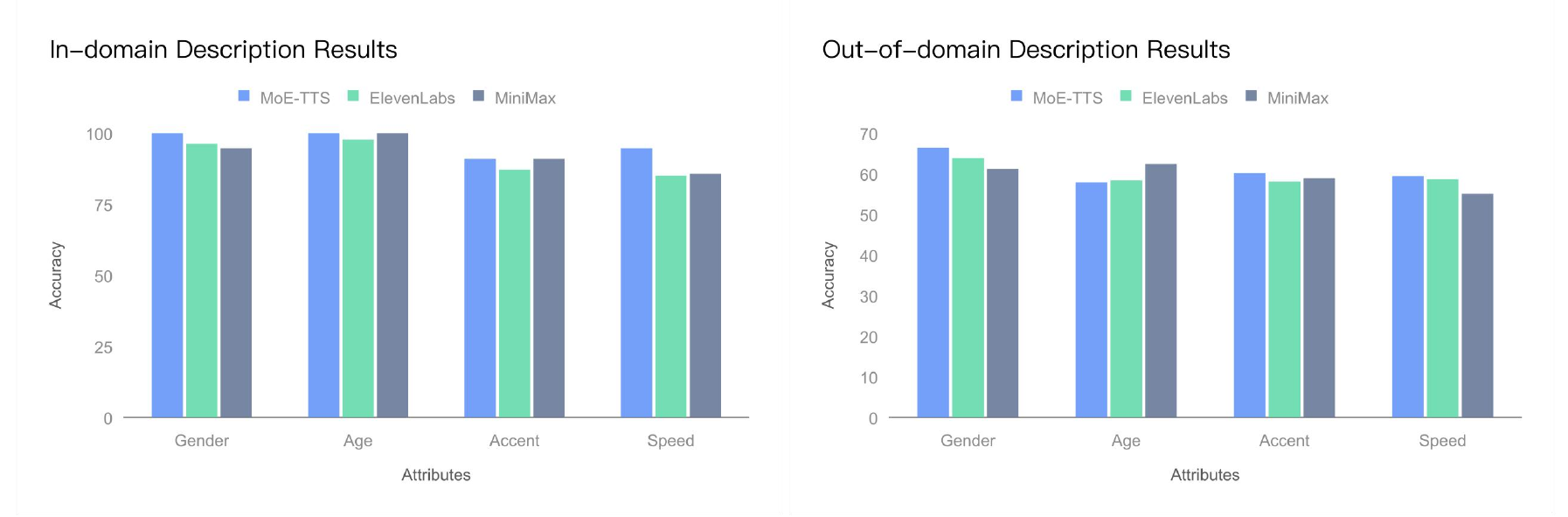}
    {
    \caption{
    We compare the accuracy of MoE-TTS, ElevenLabs, and Minimax across four fundamental voice attributes—gender, age, accent, and speed—using both in-domain and out-of-domain description test sets.
    \label{fig:fig2}
    }
    }
\end{figure*}

\paragraph{Comparison with State-of-the-Art Models}
Our evaluation compares MoE-TTS against leading commercial systems: ElevenLabs and MiniMax. We synthesized test samples using their latest publicly available APIs\footnote{As of early August 2025, the ElevenLabs multilingual\_v2 API was used, as the alpha v3 remained inaccessible.} to benchmark capabilities at the time of evaluation. Table~\ref{tab:mos} presents MOS results across test sets. For basic TTS capabilities, MoE-TTS achieves performance comparable to both commercial systems across all test sets. ElevenLabs attained the highest scores in WSSA and PA, while MiniMax excelled in SQ. Given that MoE-TTS foundational TTS capabilities were trained exclusively on open-source datasets, these results validate its potential as a novel paradigm for foundational TTS modeling. Crucially, in description-alignment dimensions (SEA and OA), MoE-TTS significantly outperforms both closed-source commercial competitors. This demonstrates that  MoE-TTS effectively leverages textual LLM pre-training knowledge to:  1) Excel on in-domain descriptions.  2) Achieve superior alignment on out-of-domain descriptions. Regarding overall scores:  MiniMax achieved the highest scores on in-domain descriptions while MoE-TTS led on out-of-domain descriptions. To analyze alignment superiority, Figure~\ref{fig:fig2} compares attribute accuracy across systems. MoE-TTS achieved the highest accuracy across most voice attributes (gender, accent, speed) in both test sets, except for the age attribute in out-of-domain descriptions where it trailed the commercial systems. Notably, all systems exhibited significant accuracy degradation on out-of-domain descriptions, highlighting inherent challenges in this domain.

\section{Limitations} 
Due to its reliance on open-source data, the current MoE-TTS implementation supports only English text descriptions as input. However, the framework shows strong potential for multilingual extension. GPU resource constraints limited our ability to evaluate MoE-TTS effectiveness across diverse LLM architectures. Key open questions include the performance impact of reduced model parameters and the scalability benefits of increased parameter counts. We defer these investigations to future work.

\section{Conclusion}
In this paper, we focus on the challenging issues posed by out-of-domain descriptions in description-based text-to-speech~(TTS) tasks. Building on the core idea of enhancing description-based TTS through the pre-trained knowledge and text understanding capabilities of large language models, we propose MoE-TTS. Our approach employs a mixture-of-experts framework that utilizes a modality-based routing strategy and modality-aware Transformer components to bridge large language models and TTS models. Experimental results demonstrate that MoE-TTS outperforms state-of-the-art closed-source commercial models using only open-source datasets.

\bibliographystyle{unsrtnat}
\bibliography{references}

\end{document}